\documentclass[pdflatex,showpacs,preprintnumbers,amsmath,amssymb,floatfix]{revtex4}
\usepackage{graphicx}
\usepackage{dcolumn}
\usepackage{bm}
\usepackage{placeins}

\usepackage{color}

\newcommand{\be}{\begin{equation}}
\newcommand{\ee}{\end{equation}}

\begin{document}

\title{Evolution of a double-front Rayleigh-Taylor system using a GPU-based high resolution thermal Lattice-Boltzmann model \footnote{Final version, postprint accepeted for publication on:  Phys. Rev. E $\bold{89}$ 043022 (2014)}}

\author{P. Ripesi$^{1*}$, L. Biferale$^1$, S. F. Schifano$^2$ and R. Tripiccione$^3$}
\affiliation{$^1$ Department of Physics and INFN, University of Tor Vergata, Via della Ricerca Scientifica 1, 00133 Roma, Italy\\
$^2$ Dipartimento di Matematica e Informatica, Universit\`a di Ferrara and INFN, Via G. Saragat 1, 44100 Ferrara, Italy\\
$^3$ Dipartimento di Fisica e Scienze della Terra, Universit\`a di Ferrara and INFN, Via G. Saragat 1, 44100 Ferrara, Italy}


\begin{abstract} 
We study the turbulent evolution originated from a system subjected to a Rayleigh-Taylor instability with a double density at high resolution in a 2 dimensional geometry using a highly optimized thermal Lattice Boltzmann code for GPUs. The novelty of our investigation stems from the initial condition, given by the superposition of three layers with three different densities, leading
to the development of two Rayleigh-Taylor fronts that expand upward and downward and collide in the middle of the cell. 
By using high resolution numerical data we highlight the effects induced by the  collision of the two turbulent fronts in the 
long time asymptotic regime. We also provide details on the optimized Lattice-Boltzmann code that we have run on a cluster of GPUs.\\
\end{abstract}

\maketitle

\section{Introduction}
\begin{center}
\begin{figure}[h!]
\includegraphics[scale=0.4,angle=90]{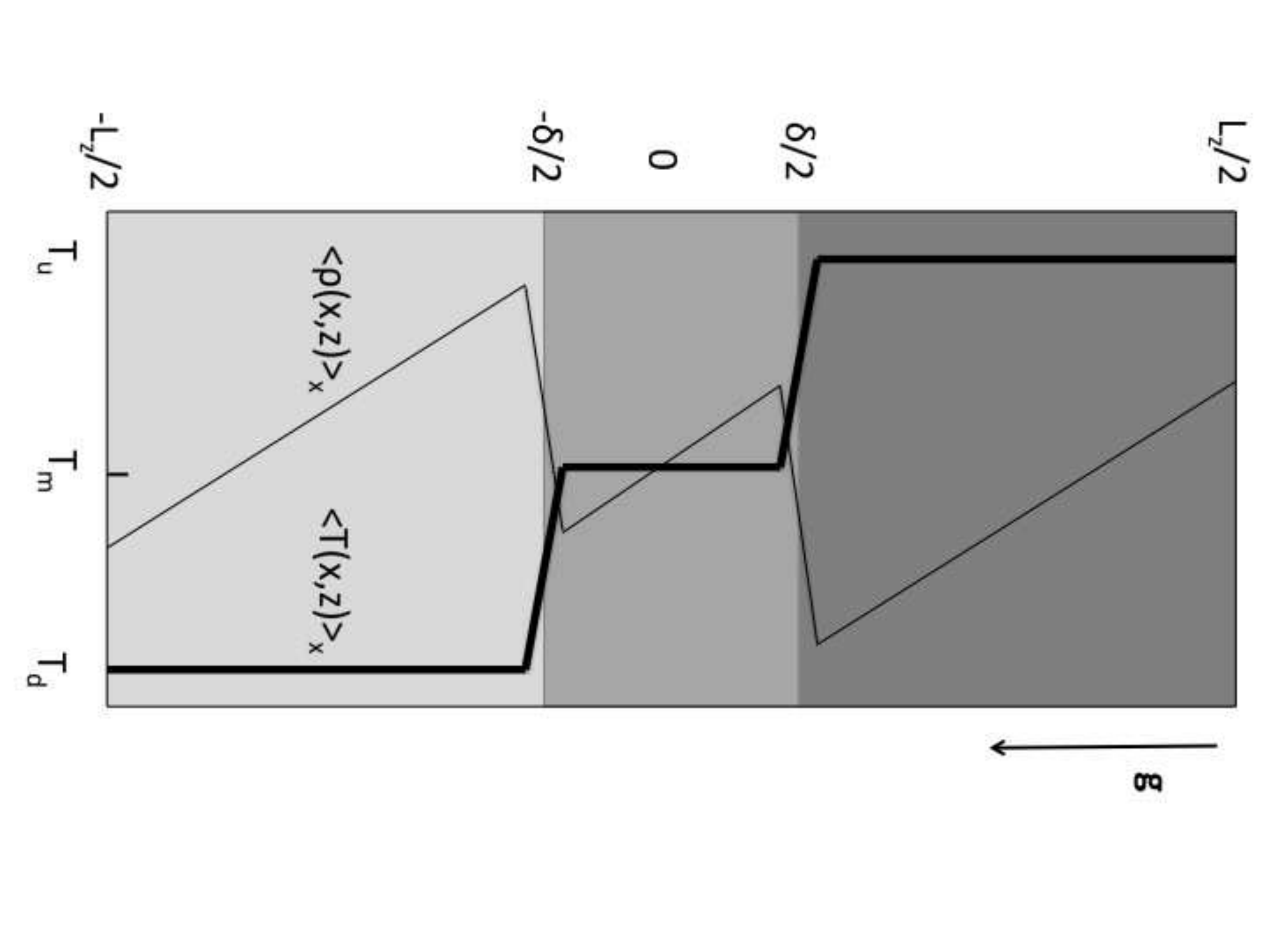}
\caption{Sketch of the initial configuration for the triple temperature Rayleigh-Taylor system. Temperature in the three regions is chosen constant, while density follow an hydrostatic profile (eq.(\ref{seteq})). The temperature jump at the interface is smoothed by a tanh profile of the order of ten grid points. The bold and thin solid lines represent the temperature and density profiles, respectively.}
\label{figure1}
\end{figure}
\end{center}

A Rayleigh-Taylor (RT) system is composed by the superposition of two layers of a single-phase fluid, with the lower lighter than the upper one and subject to an external gravity field $g>0$. In this system, the two layers mix together until the fluid reaches ``equilibrium'', characterized by a completely homogeneous environment with  hydrostatic  density/temperature profiles. Applications span a wide range of fields, such as astrophysics \cite{Cabot}, quantum physics related to the inmiscible Bose-Einstein condensates \cite{Sasaki,Kobyakov}, ocean and atmospheric sciences \cite{Spalart,Gutman}. Historically, the first theoretical work on the stability of a stratified fluid in a gravitational field is due to Rayleigh in 1900 \cite{Rayleigh}, followed about forty years later by Taylor's work on the growth of perturbations between two fluids with different densities \cite{Taylor}. Since then, the Rayleigh-Taylor instability has been intensively studied theoretically, experimentally and numerically (see, e.g., the  review of Dimonte {\it et al.} \cite{Dimonte}). Still, many  problems  remain open. 
The RT instability amounts to two main physics problems: the initial growth of perturbations between two layers of fluid with different densities and the mixing problem related to the penetration of the perturbation front through the static fluid.
The evolution of the mixing layer length, $L(t)$, follows a 'free-fall' temporal law, 
$L(t)=\alpha{g}(At)t^2$; $At=(\rho_1-\rho_2)/(\rho_1+\rho_2)$ is {\it Atwood number}
which takes into account the density differences between the upper $(\rho_1)$ and lower $(\rho_2)$ layers, $g$ is the acceleration of gravity and $\alpha$ is a dimensionless coefficient, the so-called {\it growth rate}.  Recent works \cite{Dalziel, Cook} have suggested that the value of $\alpha$ might depend also on the initial conditions. 
Beside the large-scale growth of the mixing layer, also small scale statistics have attracted the attention of many groups in recent years, both in 2d, 3d and  {\it quasi} 2d-3d geometries \cite{chertkov,Biferale,boffettaprl,boffettapre}. Moreover, different setups have been 
investigated including stratification \cite{Biferale2,Scagliarini,spiegel,frolich,robinson} and reaction \cite{biferaleepl,chertkovjfm}.
In spite of the progress made so far, most of the work in this area is limited to the classical case of double density fluids, while complex stratifications effects have not been extensively investigated. In this paper we want to present the study of an RT system which is slightly different from the ones present in the literature: we focus on the spatio-temporal evolution of a single component fluid when initially prepared with three different density layers in hydrostatic unstable equilibrium (see Figure \ref{figure1}).
Previous studies on triple density RT were limited to the case of one unstable and one stable layer \cite{Jacobs}, focusing mainly on the entrainment by the unstable flow inside the stable one. On the contrary, we have a fully unstable density/temperature distributions with two unstable layers and we need to take into account the nonlinear interactions of rising and falling plumes from each one of the two developing mixing layers, at difference from the case of the interaction of buoyant plumes with an interface \cite{Nokes}. Similarly, our case differs from the case of propagating fronts \cite{Bychkov,Bell,Modestov} because we do not have the extra effects induced by the deflagration velocity.
 The setup  is given by a two-dimensional ($2d$) $L_x{\times}L_z$ tank of fluid split in three sub-volumes at three different --initially homogeneous-- temperatures $T_u<T_m<T_d$, where each of them is in hydrostatic equilibrium $\partial_{z}p_{0}(z)=-g\rho_{0}(z)$. We enforce periodic boundary conditions in the horizontal direction.\\
The initial hydrostatic unstable configuration is therefore given by:
\be
\begin{split}
\begin{cases}
T_{0}(z)=T_d, \hspace{2mm} \rho_0^d(z)=\rho_{d}exp[-g(z-z_d)/T_d]  & -L_z/2<z<-{\delta}/2  \\
T_{0}(z)=T_m, \hspace{2mm} \rho_0^m(z)=\rho_{m}exp[-g(z-z_m)/T_m]  & -{\delta}/2<z<{\delta}/2  \\
T_{0}(z)=T_u, \hspace{2mm} \rho_0^u(z)=\rho_{u}exp[-g(z-z_u)/T_u]  & {\delta}/2<z<L_z/2  
\label{seteq}
\end{cases}
\end{split}
\ee
where $\delta$ is the width of the middle layer at temperature $T_m$, and $z_u, z_m, z_d$ are three parameters fixing the overall geometry.\\
Assuming that for each single domain we have $p_0(z)=T_0{\rho}_0(z)$, in order to be at equilibrium we require 
the same pressure at the interface, finding the 
following simple condition on  the above expressions 
\be
\rho_0^{d}(-\delta/2)T_{d}=\rho_0^{m}(-\delta/2)T_{m}, \hspace{2mm} \rho_0^{m}(\delta/2)T_{m}=\rho_0^{u}(\delta/2)T_{u}
\ee
Since $T_u<T_m<T_d$, we have $\rho_{u}>\rho_{m}>\rho_{d}$, ensuring that we have an unstable initial condition.\\
From a phenomenological point of view, the problem we are going to study is the interaction between two turbulent fronts (one originated from the upper density jump and one from the lower) when they come in contact and then evolve together like a single mixed front. Turbulent fluctuations are the driving force 
of the RT instability, so it is of interest to study how two turbulent flows interact. Moreover, comparison with the one-front  RT system is made in order to show 
how the presence of an ``intermediate'' well-mixed turbulent layer of fluid can alter the evolution of typical large scale quantities,
such as the mixing layer length and the velocity-temperature fluctuations.
The study is done using a  Lattice Boltzmann Thermal (LBT) scheme \cite{Biferale, Scagliarini} running on a GPU cluster, so our work is interesting also from the point of
 view of several architecture-specific optimization steps that we have applied to our computer code in order to boost its computational efficiency. 
This paper is organized as follows: in section II we present the equations of motion; in section III we describe the details of the lattice Boltzmann model (LBM) formulation and its implementation on GPUs.
Section IV presents the result of a large scale analysis and compares with the ``classical'' RT evolution. Conclusions close the paper in section V.

\section{Equations of motion}
The evolution of a compressible flow in an external gravity field is described by the following Navier-Stokes equations (double indexes are summed upon):
\be
\begin{split}
\begin{cases}
D_{t}{\rho}=-\partial_{i}{({\rho}u_i)} \\
\rho{D_{t}}u_i=-\partial_{i}P-{\rho}g \, \delta_{i,z}+\mu{\partial_{jj}u_i} \\
\rho{c_p}D_tT-D_tP=k{\partial_{ii}T}
\label{NS}
\end{cases}
\end{split}
\ee
where $D_t$ is the material derivate, $\mu$ and $k$ are molecular viscosity and thermal conductivity respectively, $c_p$ is the 
constant pressure specific heat and $\rho$, $T$, $P$ and ${\bf u}$ are the thermo-hydrodynamical fields of density, temperature, pressure and velocities, respectively.

In the limit of small compressibility, the parameters depend weakly on the local thermodynamics fields, so expanding pressure around its hydrostatic value $P=p_0+p$, with $\partial_zp_0=-g\rho$ and $p<<p_0$, and performing a small Mach number expansion we can write equations (\ref{NS}) as:

\be
\begin{split}
\label{eq:nsf}
\begin{cases}
D_{t}{\rho}=-\partial_{i}{({\rho}u_i)} \\
D_tu_i=-\partial_ip/\rho+g\theta/\tilde T \delta_{i,z}+\nu{\partial_{jj}u_i}\\
D_tT-u_z\gamma=\kappa\partial_{ii}T
\end{cases}
\end{split}
\ee
where $\tilde T$ is the mean temperature averaged on the whole volume, $\nu=\mu/\rho$ is kinematic viscosity, $\kappa=k/(c_p\rho)$ is thermal diffusivity and $\gamma=g/c_p$ is the adiabatic gradient for an ideal gas. From this approximation it is clear that only temperature fluctuations $\theta=T-\tilde T$ force the system; assuming the adiabatic gradient is 
negligible, $\gamma{\sim}0$, it is well know that starting from an unstable initial condition, as show in Figure \ref{figure1}, any small perturbation will lead to a turbulent mix between the cold and hot regions, developing along the vertical direction. If the adiabatic gradient is not negligible, the RT mixing does not proceed forever and stops when the mixing length becomes of the order of the stratification length scale, a further complexity that will not be studied here \cite{Biferale2}.

\section{Numerical Method}
\subsection{Thermal Kinetic Model}
In this section, we recall the main features of the lattice Boltzmann model (LBM) employed in the numerical simulations; for full details we refer the reader to the works of \cite{Philippi,Sbragaglia,Scagliarini}. 

For an ideal isothermal fluid, LBM \cite{gladrow,benzi,chen} can be derived from the continuum Boltzmann equation  in the BGK approximation \cite{bhatnagar}, upon expansion in Hermite velocity space of the single particle distribution function (PDF) $f(\bm{x},\zeta,t)$, which describes the probability to find a particle at space-time location $(\bm{x},t)$ 
and with velocity $\zeta$ \cite{he,martys,shan}. Discretization on the lattice is enforced taking a discrete finite set of velocities $\zeta{\in}[\bm{c}_1,\bm{c}_2,...,\bm{c}_M]$, where the total number $M$ is determined by the embedding spatial dimension and the required degree of isotropy \cite{gladrow}. As a result, the dynamical evolution on a discretized spatial and temporal lattice is described by a set of populations $f_l(\bm{x},t)$ with $l=1,...,M$.

\begin{center}
\begin{figure}[h!]
\includegraphics[scale=0.4,angle=0]{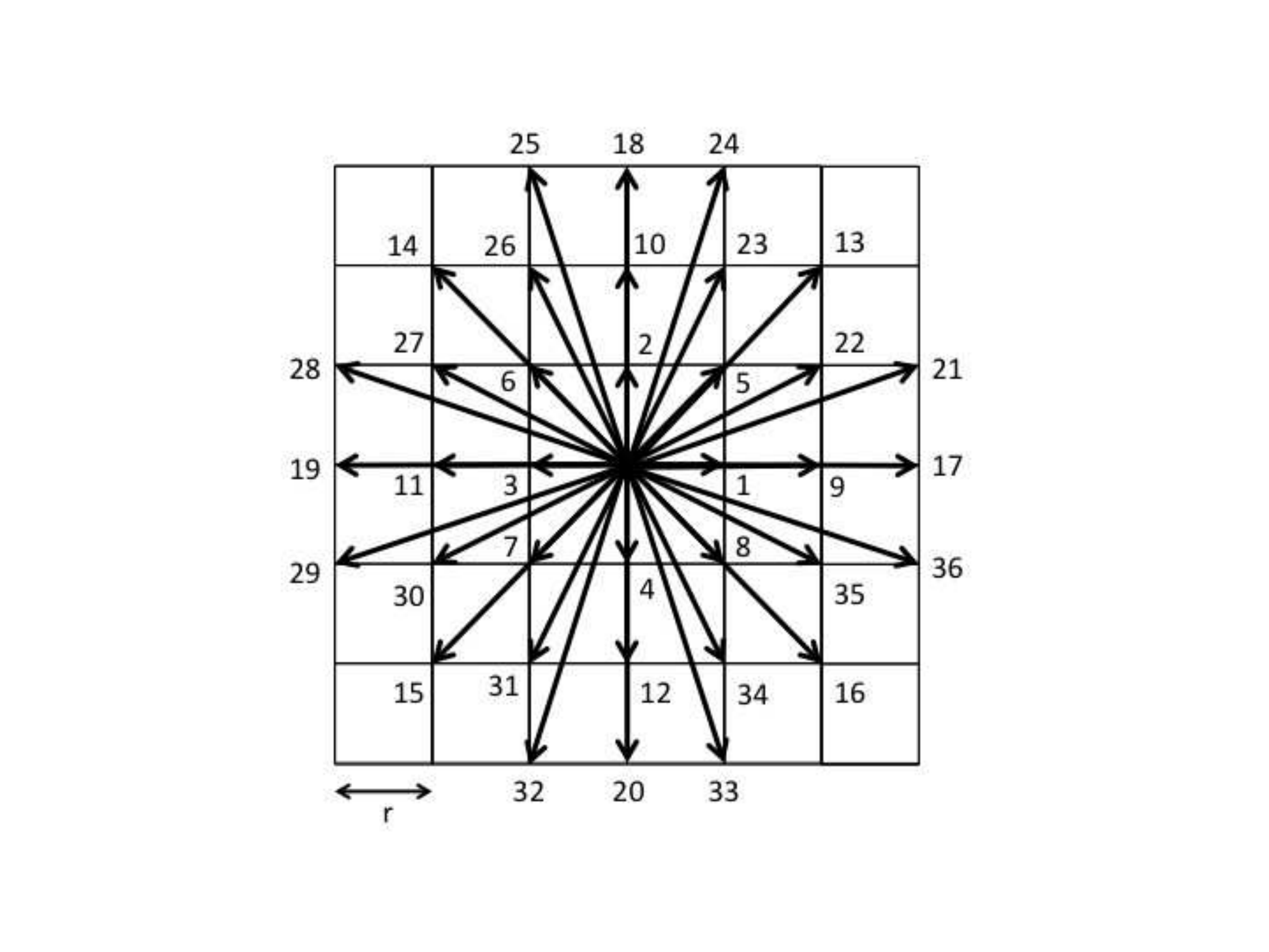}
\hspace{-8mm}
\caption{Scheme of the discretized set of $37$ velocities $c_l$ that are used by our LBM to recover the hydrodynamical behavior in the long wavenumber limit. $r\sim{1.1969}$ is the lattice constant \cite{Scagliarini}.}
\label{D2Q37}
\end{figure}
\end{center}

In the papers \cite{Philippi,Sbragaglia} it was shown that in two dimension with only one set of kinetic populations we need $M=37$ fields (the so-called {\it D2Q37} model) to recover in the Chapman-Enskog limit the continuum thermal-hydrodynamical evolution given by eq.(\ref{NS}). 
The set of speeds are shown in Figure \ref{D2Q37}, while the discretized LBM evolution is given by
\be
f_{l}({\bm x}+{\bm c}_l\Delta{t},t+\Delta{t})-f_{l}({\bm x},t)=-\frac{\Delta{t}}{\tau_{LB}}[f_{l}({\bm x},t)-f^{(eq)}_{l}].
\label{LBE}
\ee
The left-hand side of eq.(\ref{LBE}) stands for the streaming step of $f_l$, while the right-hand side 
represents the relaxation toward a local Maxwellian distribution function $f^{(eq)}_{l}$, with the characteristic time $\tau_{LB}$. 
A novelty introduced by the mentioned algorithm is that the equilibrium distribution function directly depends on the coarse grained variables plus a shift due to the local body force term \cite{Scagliarini,Philippi}:
\be
f^{(eq)}_l=f^{(eq)}_l\left[\rho,{\bf u}+{\bf g}\tau_{LB}, T+\frac{\tau_{LB}(\Delta{t}-\tau_{LB})}{d}g^2 \right ];
\ee
the macroscopic fields are defined in terms of the lattice Boltzmann populations as follows:
\be
\rho=\sum_{l}{f_{l}}, \hspace{0.2cm} {\rho}\,{\bold{u}}=\sum_{l}{\bold{c}_{l}f_{l}}, \hspace{0.2cm} d\,{\rho}\,T=\sum_{l}{\vert}{\bold{c}_{l}-\bold{u}}{\vert}^{2}f_{l}
\ee
( $d$ is the space dimensionality).
In \cite{Scagliarini,Sbragaglia}, it was shown that in order to avoid spurious terms due to lattice discretization and to recover the correct hydrodynamical description from the discretized lattice Boltzmann variables, momentum and temperature must be renormalized. This can be obtained by taking for momentum and temperature the following expressions: 
$$
\bold{u}^{(H)}=\bold{u}+\frac{\Delta{t}}{2}\bold{g}; \qquad 
T^{(H)}=T+\frac{(\Delta{t})^2g^2}{4D}.
$$
Using these renormalized hydrodynamical fields for a 2d geometry, it is known that is possible to recover the standard thermo-hydrodynamical equations (\ref{eq:nsf}) through a Chapman-Enskog expansion \cite{Scagliarini,Sbragaglia}.

\subsection{GPU optimized algorithm}
We have optimized a code that implements the LBM described above, taking into account both performance on one GPU and scaling on a fairly large number of GPUs;  our runs have been performed on a cluster based on NVIDIA C2050/C2070 GPUs. 
GPUs have a large number of small processing elements working in parallel and performing the same sequence of operation; in CUDA, 
the programming language that we have used throughout, each sequence of instructions operating on different data is called a {\em thread}.
Optimization focuses along three lines: i) organizing data in such a way that it can be quickly moved between GPU and memory; ii) ensuring that a large number of GPU-threads operate independently on different data items with as little interference as possible, and iii) organizing data moves among GPUs minimizing the idle time of each GPU as it waits for data from another node.\\
Concerning data organization, we first split a lattice of size $L_x \times L_z$
on $N_p$ GPUs along the $x$ dimension; each GPU handles a sublattice of $L_x/N_p
\times L_z$ points. Lattice data is stored in memory in column-major order and 
we keep in memory two copies of the lattice: at each time step, the code reads
from one copy and writes to the other. This choice requires more memory, but it allows to map one GPU thread per lattice site and then
process all threads in parallel. Arrays of LB populations are stored in memory
one after the other (this is usually referred to as Structure-of-Arrays [SOA]);
this scheme helps {\em coalescing} memory accesses to different
lattice sites that the GPU processes in parallel, and helps increase bandwidth.
\begin{center}
\begin{figure}
\includegraphics[width=0.6\textwidth]{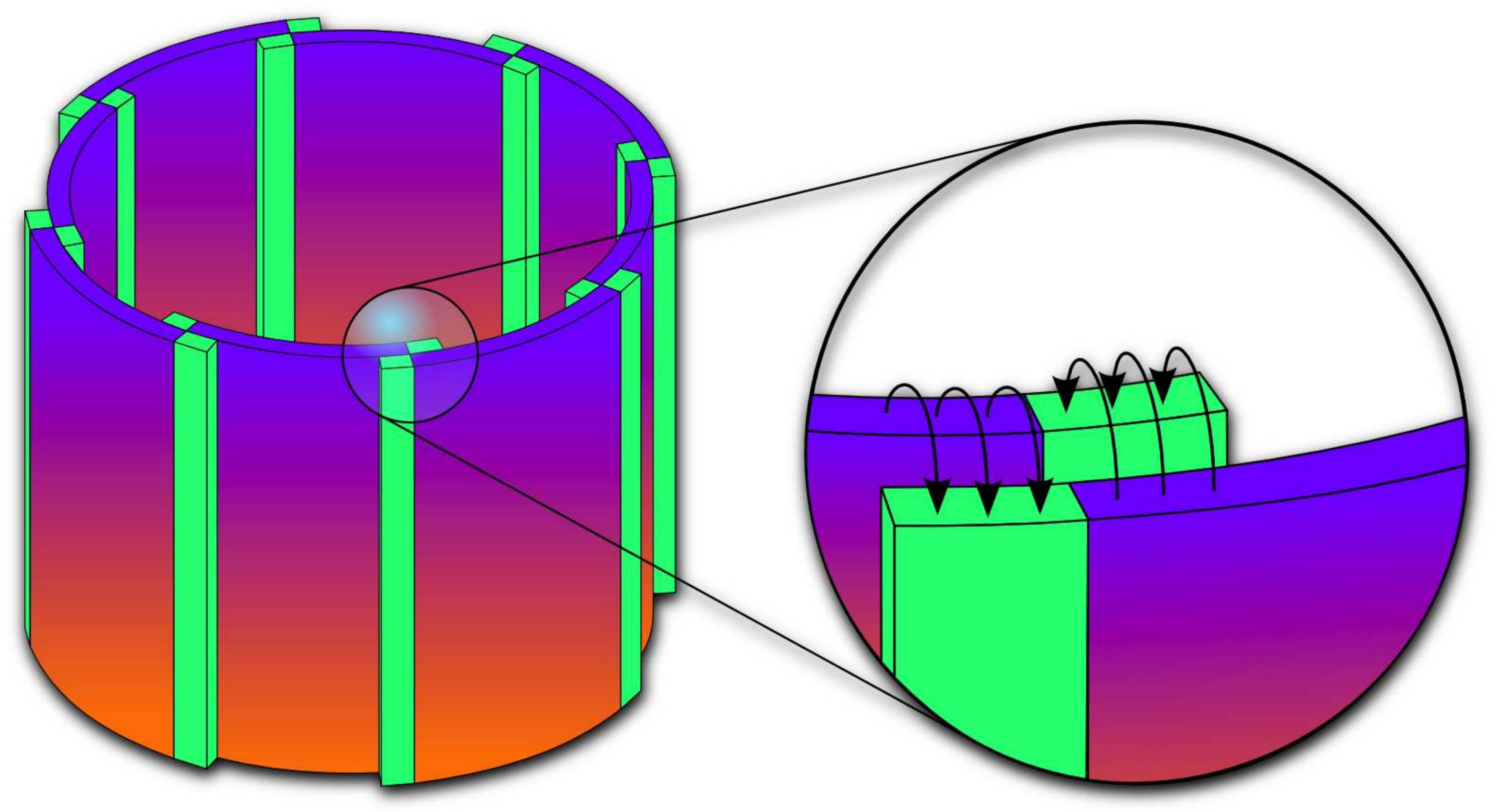}
\caption{(Color online) Tiling of the physical lattice with periodic boundary conditions in the $x$ direction on several processing nodes; the picture shows the halo columns that contain copies of the lattice points processed by the neighboring processors.}
\label{fig:lattice}
\end{figure}
\end{center}
The physical lattice is surrounded by halo-columns and rows, see Figure
\ref{fig:lattice}. For a physical lattice of size $L_x \times L_z$, we allocate
a grid of $NX \times NZ$ lattice points, $NX = H_x +L_x +H_x$, and $NZ = H_z
+L_z +H_z$. $H_x, H_z$ are the sizes of the halos used to establish data continuity between GPUs working on
adjoining sublattices and to enforce periodic boundary conditions in the $x$
direction. This makes the computation uniform for all sites, so we avoid
thread divergences, badly impacting on performance. We have $H_x = 3$,
and $H_z = 16$; the halo in $Z$ is larger than needed by the algorithm, in order to keep data aligned (data words must be allocated in multiples of 32), and to maintain cache-line alignment in multiples of 128 Bytes.\\
As customary for GPUs, the host starts the execution of each time step,
corresponding to four main kernels: first the {\tt periodic boundary conditions} step exchanges columns
of the z-halos, then three steps follow that implement (i)  the free propagation {\tt (propagate)} expressed by the lhs of (\ref{LBE}), (ii) 
the rigid boundary conditions {\tt (bc)} at the top and bottom walls and (iii) the collisions {\tt (collide)} in the rhs of (\ref{LBE}). 
Step {\tt propagate} moves each population of each site to a different lattice
site, according to its velocity vector. Computerwise this corresponds to memory
accesses to sparse addresses. Two options are available: {\em push} moves all
populations from the {\em same} site to their appropriate destinations; while
{\em pull} gathers populations from neighbor sites to the {\em same}
destination; in the first case one has aligned memory reads and misaligned
writes, while the opposite is true in the second case; we have tested both
options and then settled for {\em pull} that offers $\simeq 20\%$ higher
bandwidth.
Step {\tt bc} executes after {\tt propagate} in order to enforce boundary conditions at the top and bottom of the cell; it adjusts population values at sites with coordinates $z=0,1,2$ and $z=L_z -3, L_z-2$ and $L_z-1$; its computational impact is fully negligible, so we do not apply significant optimization steps here.
Finally, {\tt collide} performs the collision phase of the LB procedure. Step {\tt collide} executes in parallel on a large number of threads: no
synchronization is necessary  because all threads read data
from one copy of the lattice (the {\tt prv} array) and write to the other copy
(the {\tt nxt} array); {\tt prv} and {\tt nxt} swap their role at the following
time step. Some care is needed to find the optimal number of threads: if this
value is too small, available computing resources are wasted, while if it is too
large there is not enough space to keep all intermediate data items on
GPU-registers and performance drops quickly. For each data points {\tt collide}
executes $\approx 7600$ double precision floating-point operations
(some of which can be optimized away by the compiler); $\approx
72\%$ of them can be cast as FMAs (fused multiply add), in which the operation
$a \times b +c$ is performed in just one step.\\

\begin{table}[b]
\begin{tabular}{l|r|rr|rr}
\toprule
-                         & QPACE & 2-WS  & C2050 & 2-SB  & K20X        \\
\hline
P (GFlops)                & 15    & 60    & 172   & 166   & 412         \\
MLUPS                     & 1.9   & 7.7   & 22    & 21.7  & 124         \\
E/site ($\mu J$/site)     & 56    & 34    & 10    & 12    & 4.2         \\  \hline
\label{compare}
\end{tabular}
\caption{
Performance comparison of our LB code among several architectures, based
on the results of \cite{iccs10,parcfd12,parcfd11}
2-WS (2-SB) are Intel dual-processor systems based on the Westmere (Sandy
Bridge) micro-architecture; C2050 (K20X) are NVIDIA GPUs based on the
Fermi (Kepler) processors.
}
\end{table}

\begin{center}
\begin{figure}[b]
\includegraphics[width=0.6\textwidth]{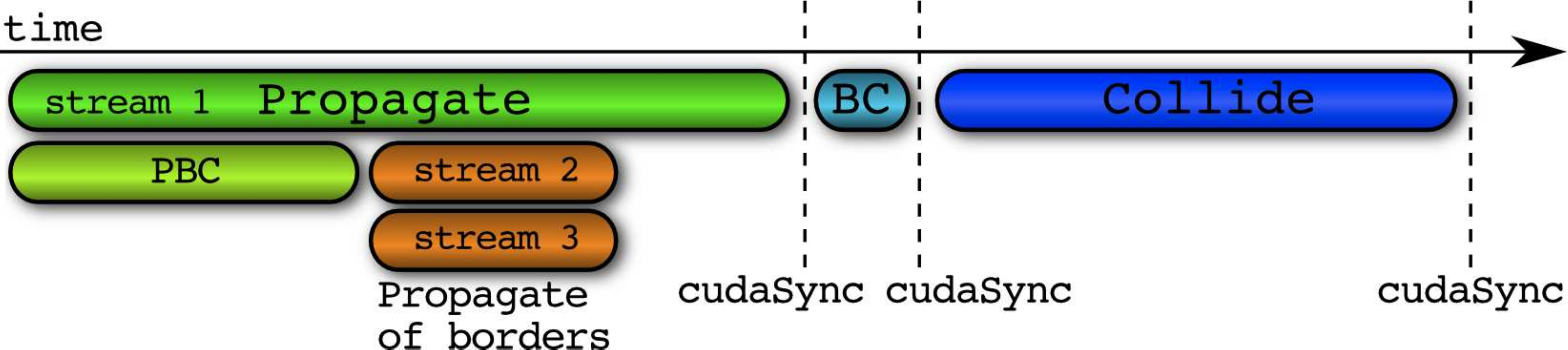}
\caption{(Color online) Time schedule of the main step of the complete program; the picture shows that the time associated to data transfer between processing elements can be overlapped with computation.}
\label{fig:timeSequence}
\end{figure}
\end{center}

We now consider the optimization steps for data exchange between different GPUs,
i.e., for step {\tt pbc} of the program.  In principle one first moves data from a strip of lattice points of one GPU to the halo region of the logically neighboring GPU, and then executes the streaming phase of the program (i.e. the {\tt propagate} function): this means that one has to wait till data transfer has completed. They key remark is that fresh halo data is only needed when applying {\tt propagate} to a small number of lattice sites (close to the halos). We then divide {\tt propagate} in three concurrent {\em streams} (in CUDA jargon). One stream handles most lattice points (those far away from the halos), while the two remaining streams handle GPU-to-GPU communications followed by {\tt propagate} on the data strips close to the halos (one process handles the halo columns at right and the other those at left). In this way the complete program follows the time pattern shown in Figure \ref{fig:timeSequence}, and effectively hides almost all communication overheads (up to $32$ GPUs, the size of the machine that we have used for our runs). 
Over the years we have developed several version of this code optimized for a number of HPC systems. As early as 2010 we had a version \cite{iccs10,BiferalePTRS} for the QPACE massively parallel system~\cite{qpace}; more recently, we carefully optimized the code for multi-core commodity CPUs~\cite{parcfd12} and many-core GPUs \cite{parcfd11}.
Table (I) summarizes our performance results. This table cover three generations of HPC processors: the early PowerCellX8i of QPACE (2008), the Intel Westmere CPU and the NVIDIA C2050 (2010), and the Intel Sandybridge CPU and NVIDIA K20 (2013). Note that for each processor we consider a code specifically optimized for its architecture, exploiting several levels of parallelism (vector instructions and core parallelism).
Table (I) shows the sustained performance of the full code, the number
of lattice-sites updated per second (MLUPS, a user friendly figure-of-merit for
performance) and the approximate value of the energy used to update one site.
Our numbers refer to the performance of just one processing node: this is a reasonable approximation to actual performance up to $\simeq 32$ nodes, since communication overheads can be successfully hidden, as discussed above.
Table (I) shows a significant improvement of performances as newer processors appear; it is interesting to underline that -- at fixed time --  GPUs offers 2-3X better performance than multi-core CPUs, while at the same time being roughly 3X more energy efficient.

Further optimizations that we have applied to our code after the present work has been completed offer even higher performances: i.e. a system of two CPUs and two K20X GPUs now breaks the 1 Tflops barrier for single-node sustained double-precision performance, see \cite{sbacpad13}.
We conclude this section underlying that the set of optimization steps that we have applied to our 2D code are expected to be equally efficient in 3D for the code running on each processing node. In this case, memory needs are larger, so it may be necessary to use a larger number of processing node, and multi-node scaling would have to be studied carefully.

\section{Data and Analysis}
Here we present the results obtained from the numerical simulations of the 2d $L_x{\times}L_z$ RT systems putting 
at the middle of the vertical domain a layer of depth $\delta$ at an intermediate temperature $T_m=(T_u+T_d)/2$. For comparison, we also perform simulations for the usual RT configuration i.e. with $T=T_u$ on the upper half and $T=T_d>T_u$ on the lower half of the cell. In this study we limit ourselves to the case of negligible stratification. Within each layer, temperature values are chosen to set $At<<1$. Anyhow, it is important to stress that the algorithm is also applicable to strongly stratified flows \cite{BiferalePTRS}. All physical parameters are listed in Table (II).
\begin{table}
\begin{tabular}{l*{15}{c}}
\toprule
{Case} & {\em $At$\/} & {\em $At_u$\/} & {\em $At_d$\/} & {\em $ L_x$\/} & {\em $L_z$\/} & {\em $\delta$\/} & {\em g\/} & {\em $T_u$\/} & {\em $T_m$\/} & {\em $T_d$\/} & {\em $N_{conf}$\/} & {\em $L_{ad}$\/} & {\em $Ra$\/} & {\em $Re$\/} & {\em $Pr$\/}\\
\hline
single-front & 0.025  &        &             & 2400  & 6144 &      & $1.5{\times}10^{-5}$ & 0.975 &         & 1.025       & 7 & 6666 & $10^{10}$ & $6{\times}10^4$ & 1\\
double-front & 0.025  & 0.0126 & 0.0123      & 2400  & 6144 & 500  & $1.5{\times}10^{-5}$ & 0.975 & 1       & 1.025       & 7 & 6666 & $10^{10}$ & $6{\times}10^4$ & 1\\
\hline
\label{tablerun}
\end{tabular}
\caption{ Parameters for the RT runs. Total Atwood number, $At=(T_d-T_u)/(T_d+T_u)$; upper Atwood number, $At_u=(T_m-T_u)/(T_m+T_u)$; lower Atwood number, $At_d=(T_d-T_m)/(T_d+T_m)$; gravity $g$; temperature in the upper region, $T_u$; temperature in the middle region, $T_m$; temperature in the lower region; $T_d$; number of independent RT evolution, $N_{conf}$; adiabatic length $L_{ad}=2{\Delta}T/g$; maximum Rayleigh number, $Ra=\Delta{T}L_z^3g/{\nu}{\kappa}$; Reynolds number, $Re=VL_z/{\nu}$ ($V=<v_z^2>_x^{1/2}$); Prandtl number, $Pr=\nu/{\kappa}$.}
\end{table}
The initial width of the intermediate layer is chosen $\delta{\sim}L_z/10$, in order to avoid possible confining effects by the vertical boundaries during the merging of the middle temperature layer.
\begin{center}
\begin{figure*}
\includegraphics[scale=0.15,angle=0]{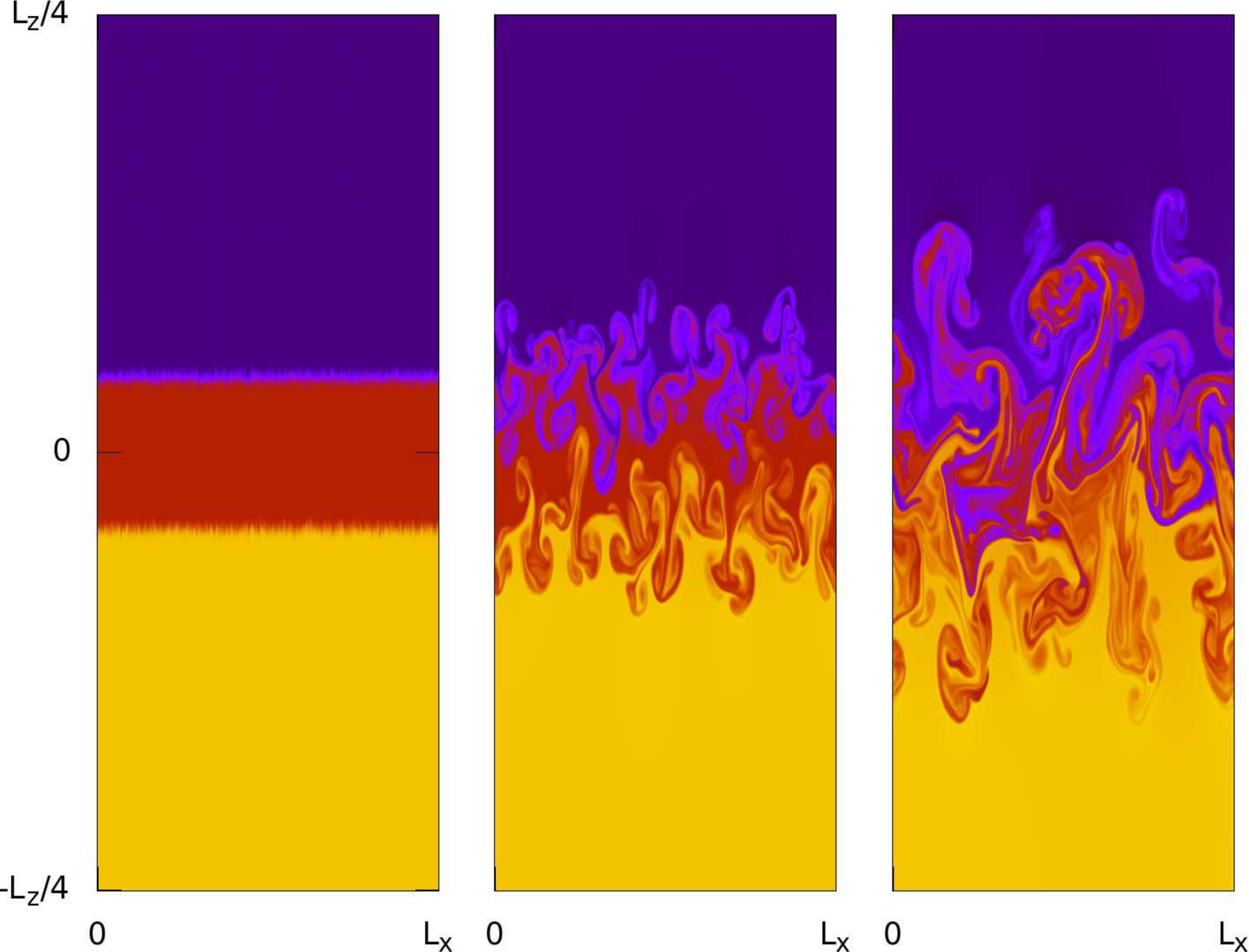}
\hspace{4mm}
\includegraphics[scale=0.15,angle=0]{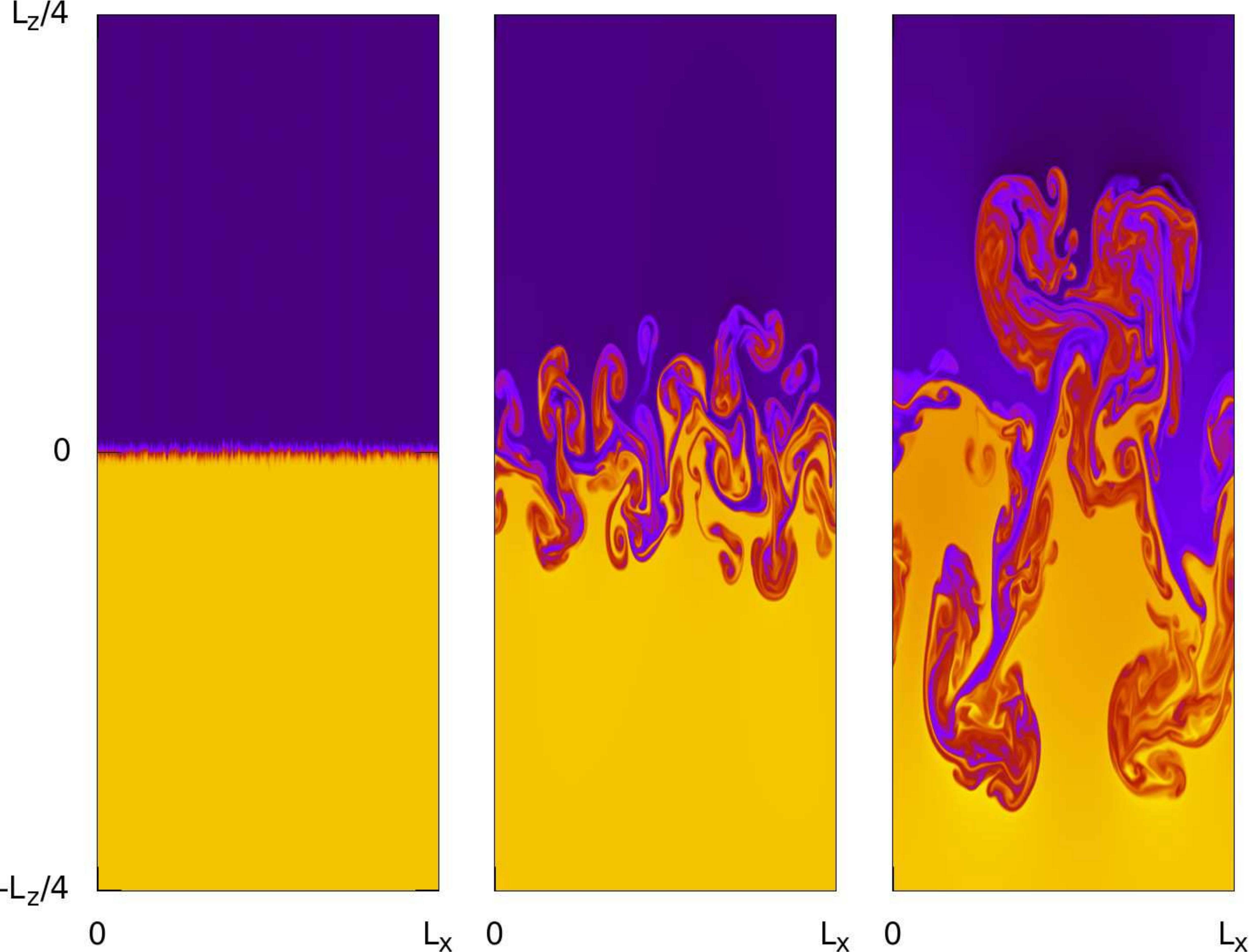}
\hspace{4mm}
\includegraphics[scale=0.15,angle=0]{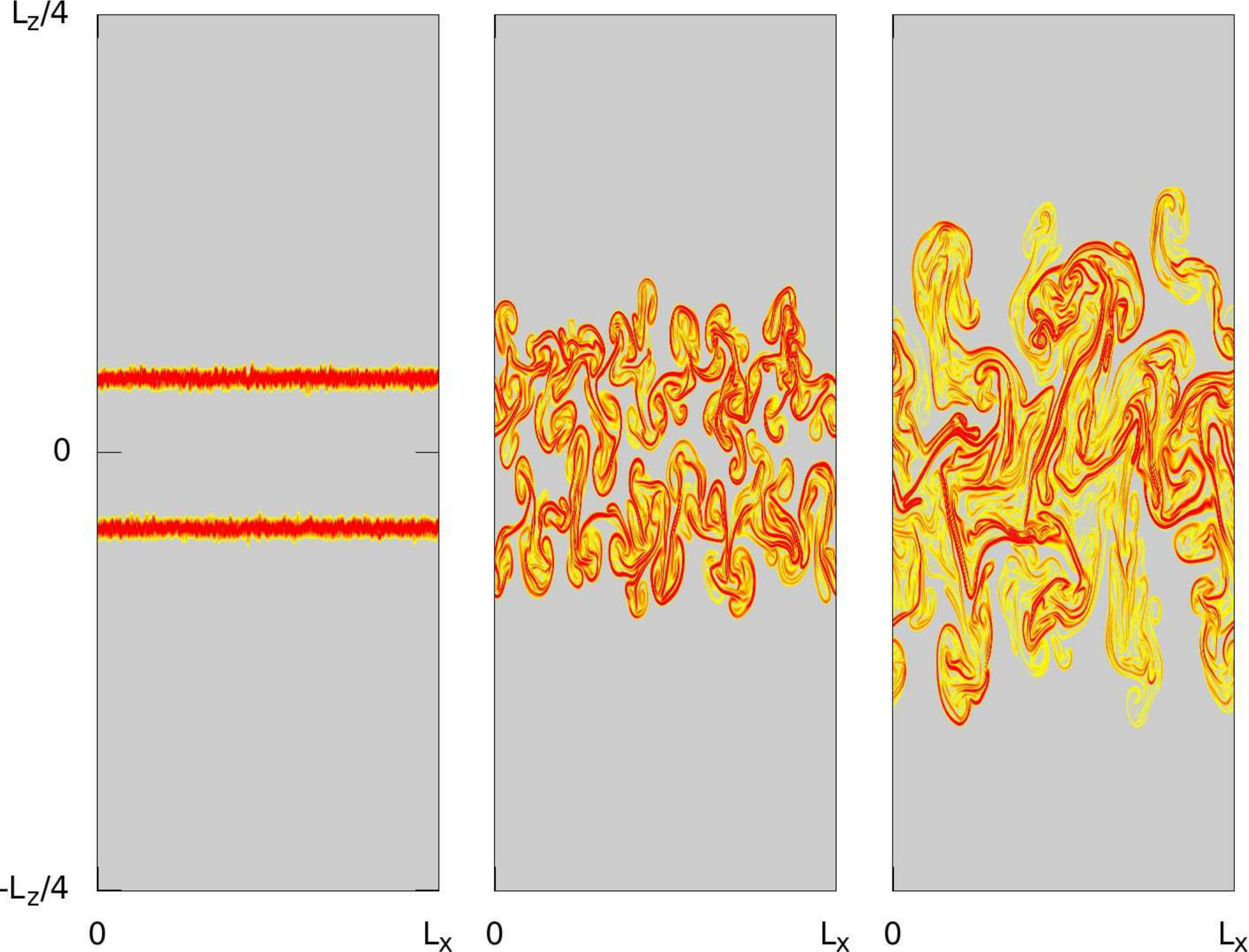}
\hspace{4mm}
\includegraphics[scale=0.15,angle=0]{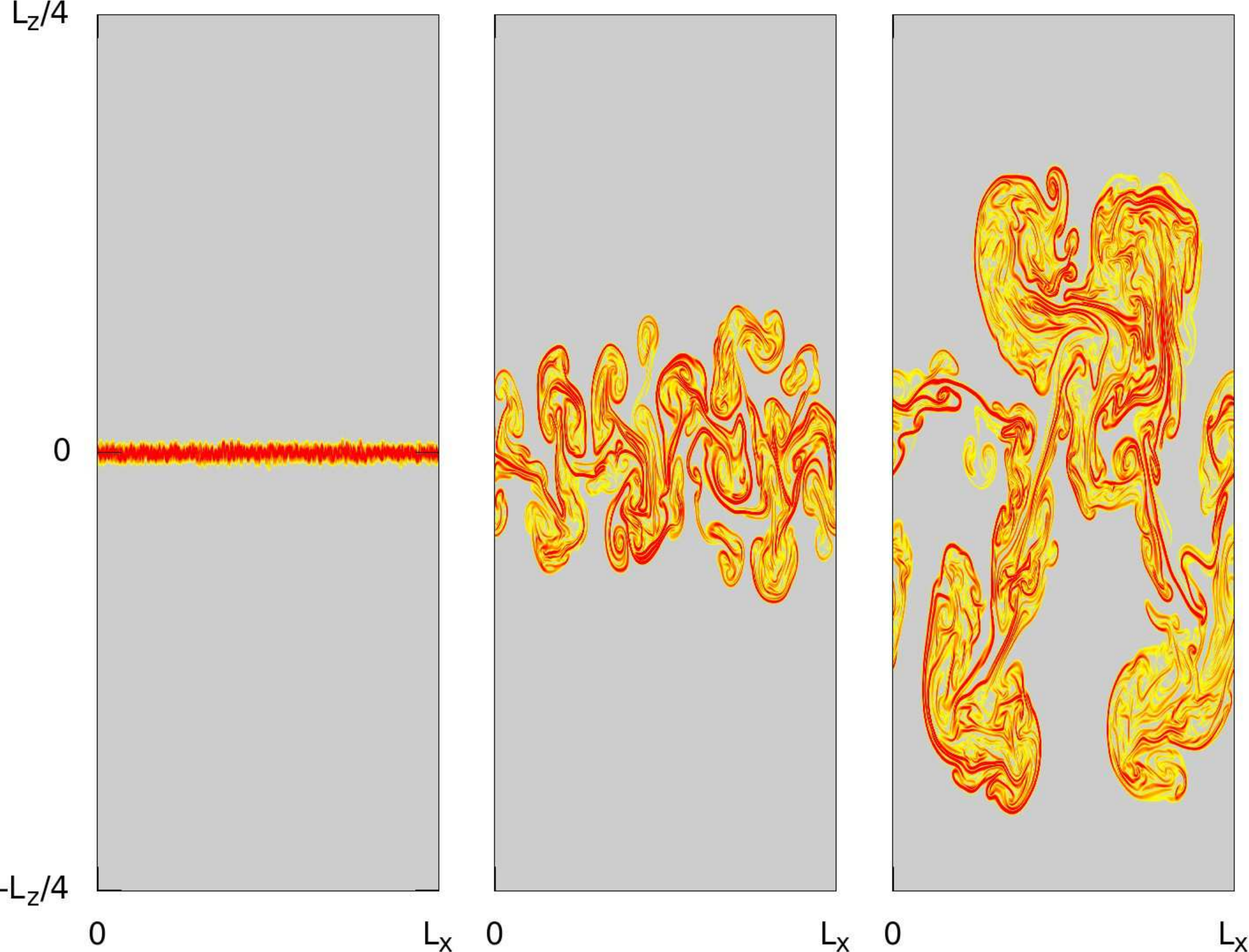}
\hspace{4mm}
\includegraphics[scale=0.15,angle=0]{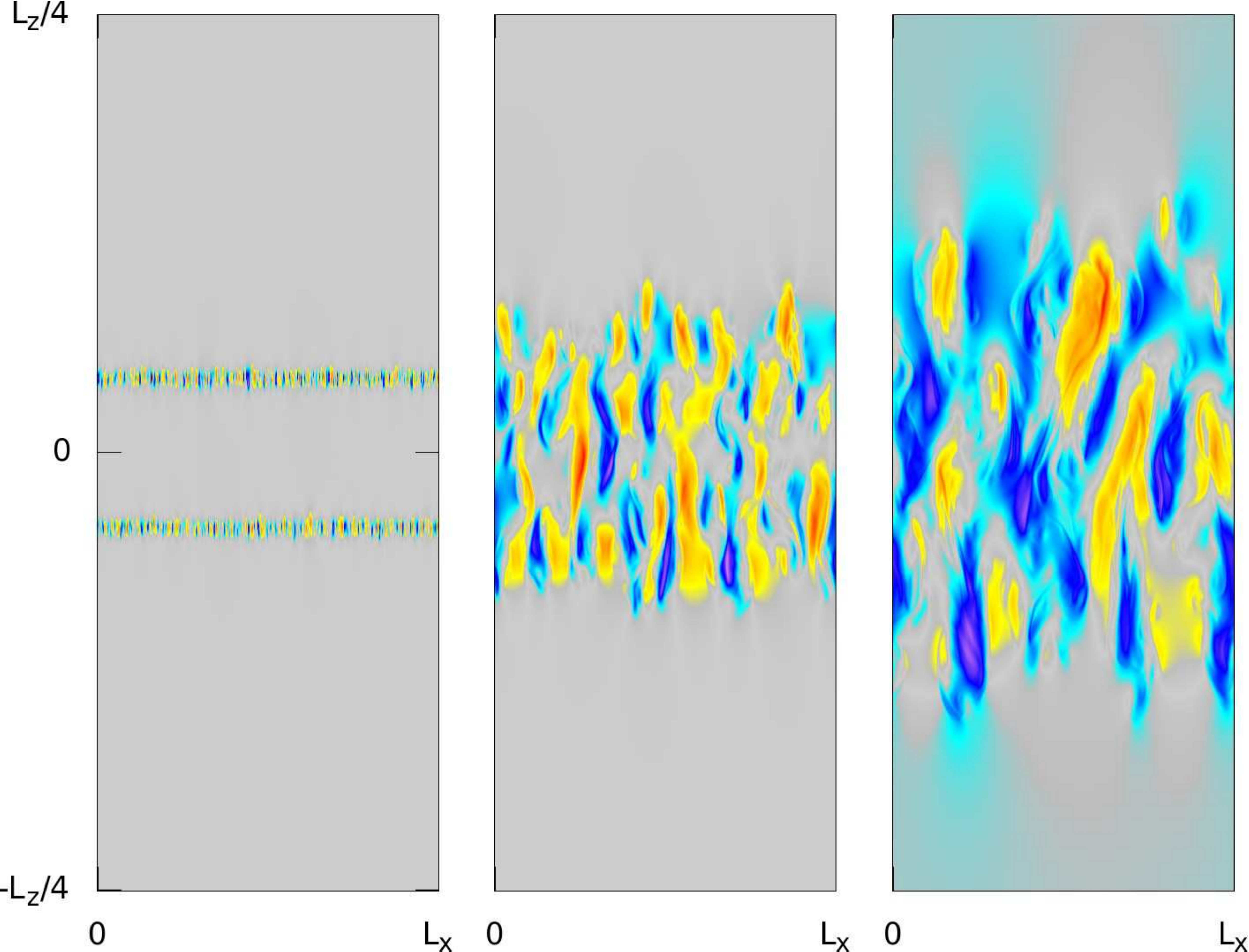}
\hspace{4mm}
\includegraphics[scale=0.15,angle=0]{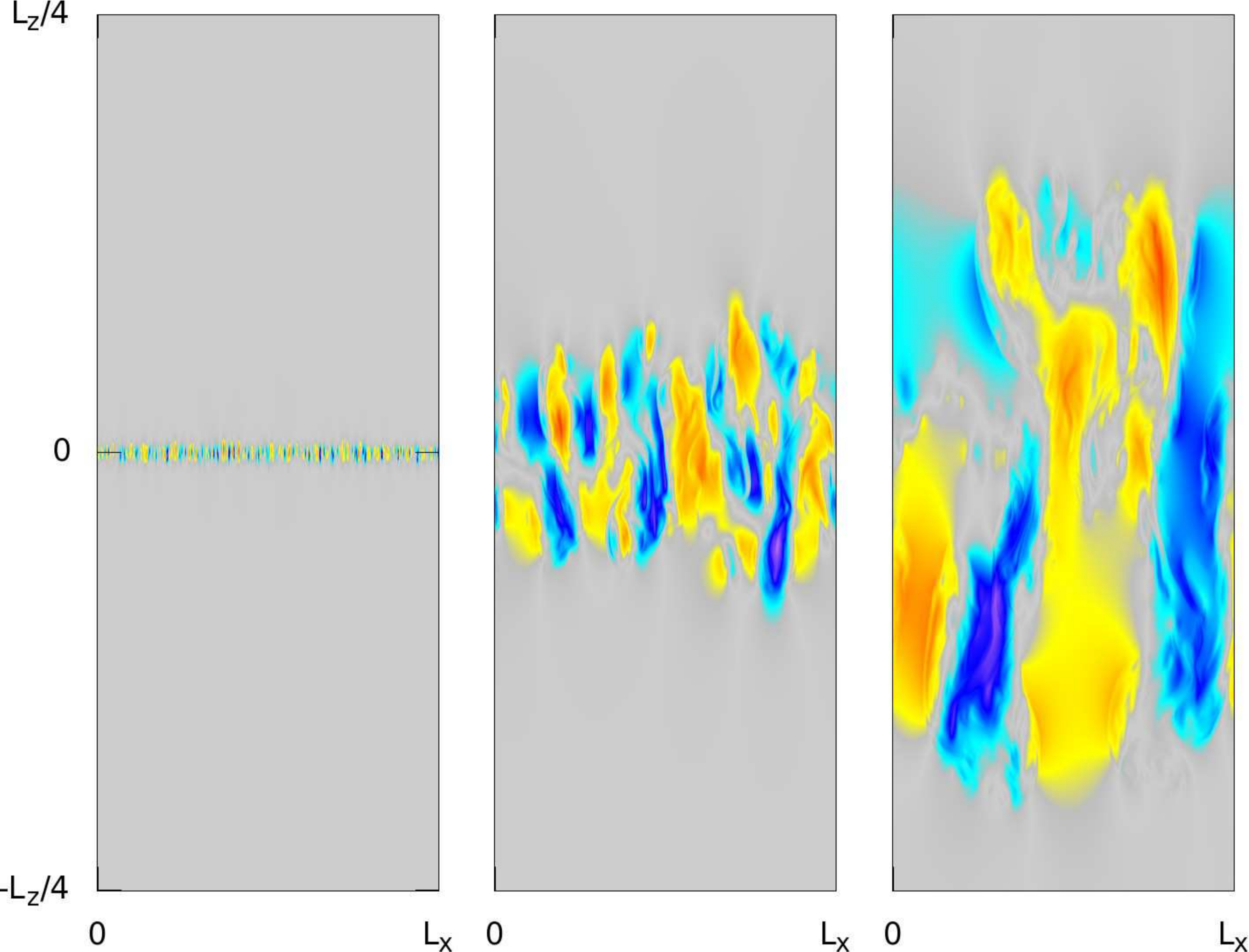}
\caption{(Color online) Snapshot of the temperature (top row), temperature gradients (middle row) and 
 vertical velocity (bottom row)  for the double-front (left) and single-front (right) RT 
at $t/\tau=0,2,3$, with $\tau=\sqrt{L_x/(gAt)}$.}
\label{figureT}
\end{figure*}
\end{center}
Snapshots of the temperature, temperature gradient and vertical velocity are shown in Figure \ref{figureT}, taken at three different time points $(t=0,2,3)\tau$ during the evolution; ${\tau}=\sqrt{L_x/(gAt)}$ is the typical RT normalization time. Vertical temperature profiles $${\overline{T}}(z)=\frac{1}{L_x}{\int dx T(x,z,t)}$$ are also shown in Figure \ref{profili1}, comparing the double-front RT with the single-front RT.

The triple density fluid starts to merge under the effect of the instability, causing the development of two fronts in correspondence of the two temperature/density interfaces. 
These two fronts continue to mix separately the  region of the flow between  $T_u$ and $T_m$  (referred as upper front) 
and the region between $T_m$ and $T_d$ (referred as lower front), until they get in touch at $t {\sim} 2.5 \tau$. At this time, the $T_m$ layer of fluid is greatly eroded and the flows with temperatures $T_u$ and $T_d$ start to interact.
Unlike the classical RT system, in this case we have also the action of turbulent viscosity exerted by one front on the other. This effect leads to a slowing down in the growth of the mixing rate of the double-front experiment with respect to the one-front case.
\begin{center}
\begin{figure}[h!]
\includegraphics[scale=0.35,angle=270]{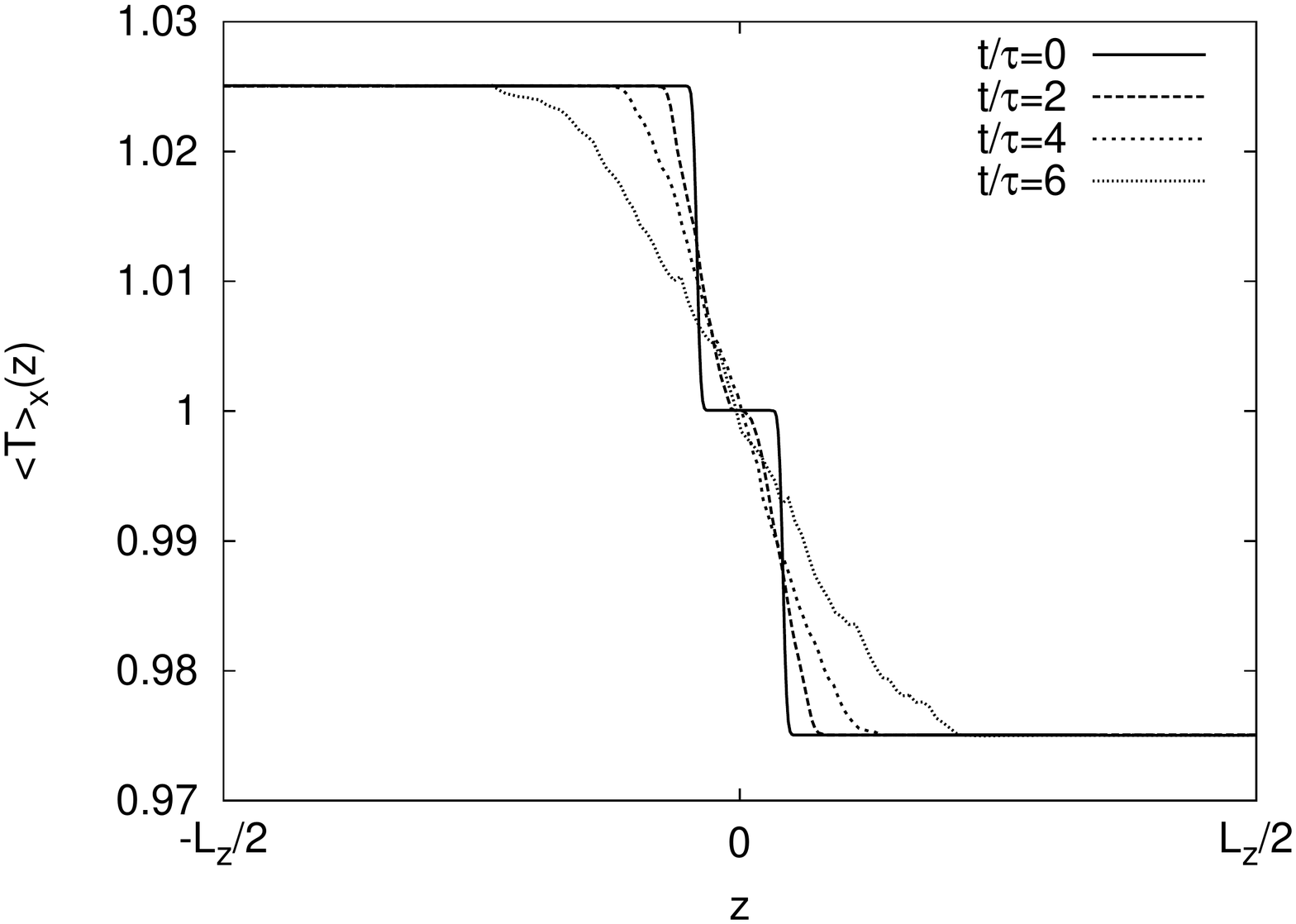}
\includegraphics[scale=0.35,angle=270]{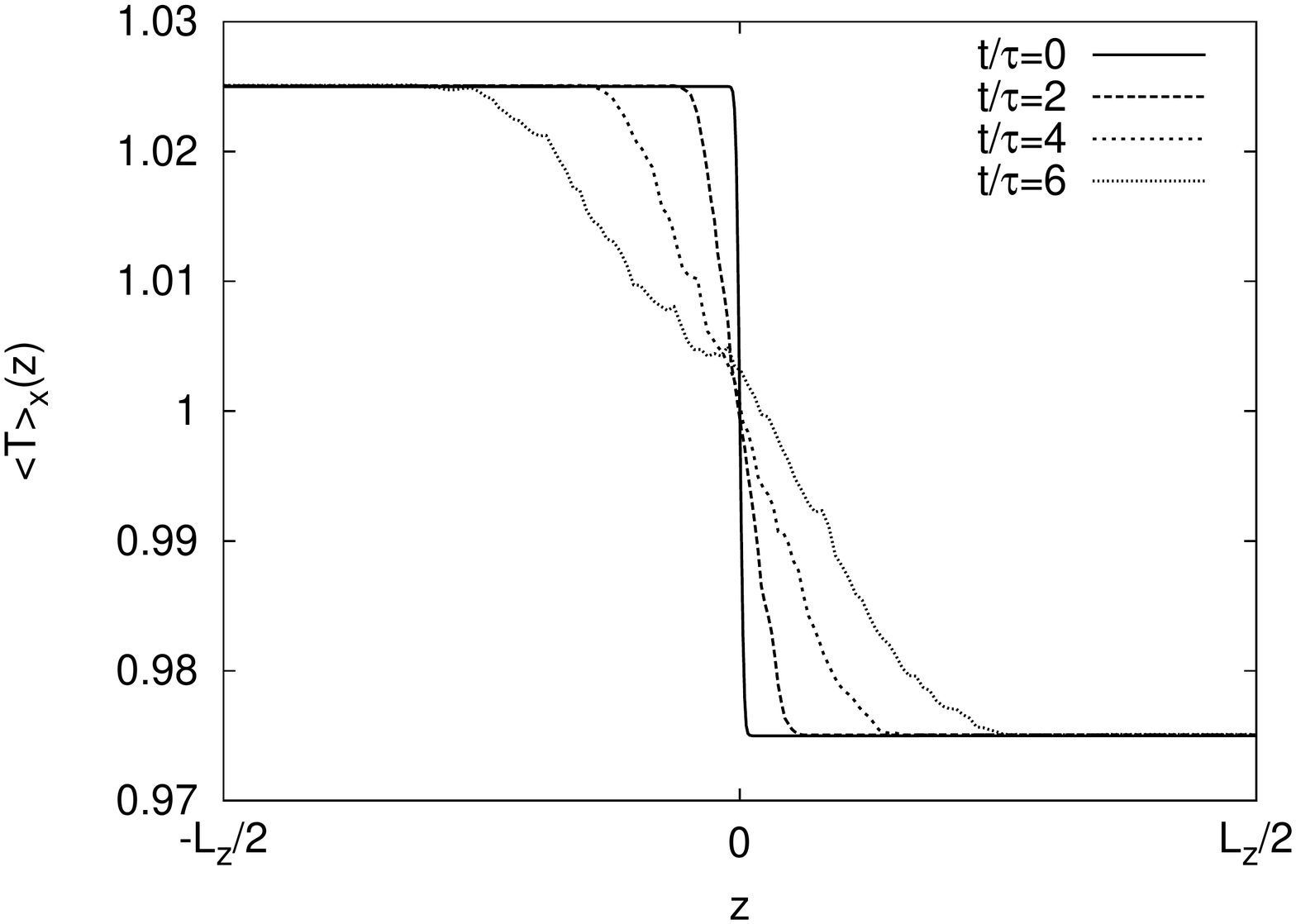}
\caption{Temperature profile for the double-step case (left) and single step case (right) at various times.}
\label{profili1}
\end{figure}
\end{center}
A useful quantity to describe the  mixing layer extension is the {\it mixing length} $L(t)$,
 defined as the region where the mean temperature profile is within a given range, e.g. 
 $\overline{T}(z) \, \in [(1+a)\,T_{u}:(1-a)\,T_{d}]$, (typically $a=0.05(T_d-T_u)$). 
An alternate way to evaluate $L(t)$ uses the following integral law \cite{Cabot}:
\be
L(t)={\int dz{\Theta}\biggl[\frac{\overline{T}(z,t)-T_u}{T_d-T_u}\biggr]}
\ee
where ${\Theta}[b]$ is a  function  with a tent map profile: 
\be
\begin{split}
\begin{cases}
{\Theta}[b]=2b & 0<b<1/2, \\
{\Theta}[b]=2(1-b) & 1/2<b<1.\\
\end{cases}
\end{split}
\ee
There is a vast literature based on observation, dimensional analysis and self-similar assumptions \cite{Read,Youngs} which shows that the mixing layer length follows a quadratic evolution in time:
\be
L(t)={\alpha} (At) g \, t^{2}
\label{eqL}
\ee
where  $\alpha$ is a dimensionless parameter named ``growth rate''. 
Taking the square of the time derivate of eq.(\ref{eqL}), we have the following self-similar scaling \cite{Ristorcelli}:
\be
[\dot{L}(t)]^2=4{\alpha}g (At) \,L(t).
\label{eqL2}
\ee
Considering our double-front problem, we have to write an expression for $L(t)$ which must take into account the different 
 nature of the system respect to the classical case. For this purpose we use two distinct mixing lengths, one for the upper front, $L_u(t)$,
 and one for the lower front, $L_d(t)$, defined as:
\be
\begin{split}
L_u(t)={\int dz{\Theta}\biggl[\frac{\overline{T}(z,t)-T_u}{T_m-T_u}\biggr]},\\
L_d(t)={\int dz{\Theta}\biggl[\frac{\overline{T}(z,t)-T_m}{T_d-T_m}\biggr]}.
\label{3L}
\end{split}
\ee
Starting from eq.(\ref{eqL}), we can write its double-front counterpart as
\be
L_u(t)={\alpha}_u(At_u)g \, t^{2},\qquad
L_d(t)={\alpha}_d(At_d)g \, t^{2}
\label{3L2}
\ee
where $At_u=(T_m-T_u)/(T_m+T_u)$ and $At_d=(T_d-T_m)/(T_d+T_m)$ are the upper and lower Atwood numbers and $\alpha_u$ and $\alpha_d$ are the upper and lower growth rate, respectively.
Within the double mixing length approach, we can define the total mixing length of the fluid as the sum of the 
upper and lower components, $L_{tot}(t)=L_u(t)+L_d(t)$.
Using the previous expression (\ref{3L2}) we have
\be
\label{eqtot}
L_{tot}(t)=\alpha_ug(At_u)\,t^2+\alpha_dg (At_d)\,t^2 \equiv {\alpha_{tot}}g (At)\,t^2
\ee
where we have introduced the definition for the total growth rate in this double layer case as $ \alpha_{tot} = (\alpha_u At_u +\alpha_d At_d)/At$
An important  advantage of eq.(\ref{eqtot}) is that it is local in time, so we may extract the coefficient ${\alpha}_{tot}$ by a simple evaluation of the plateau in the ratio $\dot{L}_{tot}^2/L_{tot}$ at each time. 
\begin{center}
\begin{figure}[h!]
\includegraphics[scale=0.35,angle=0]{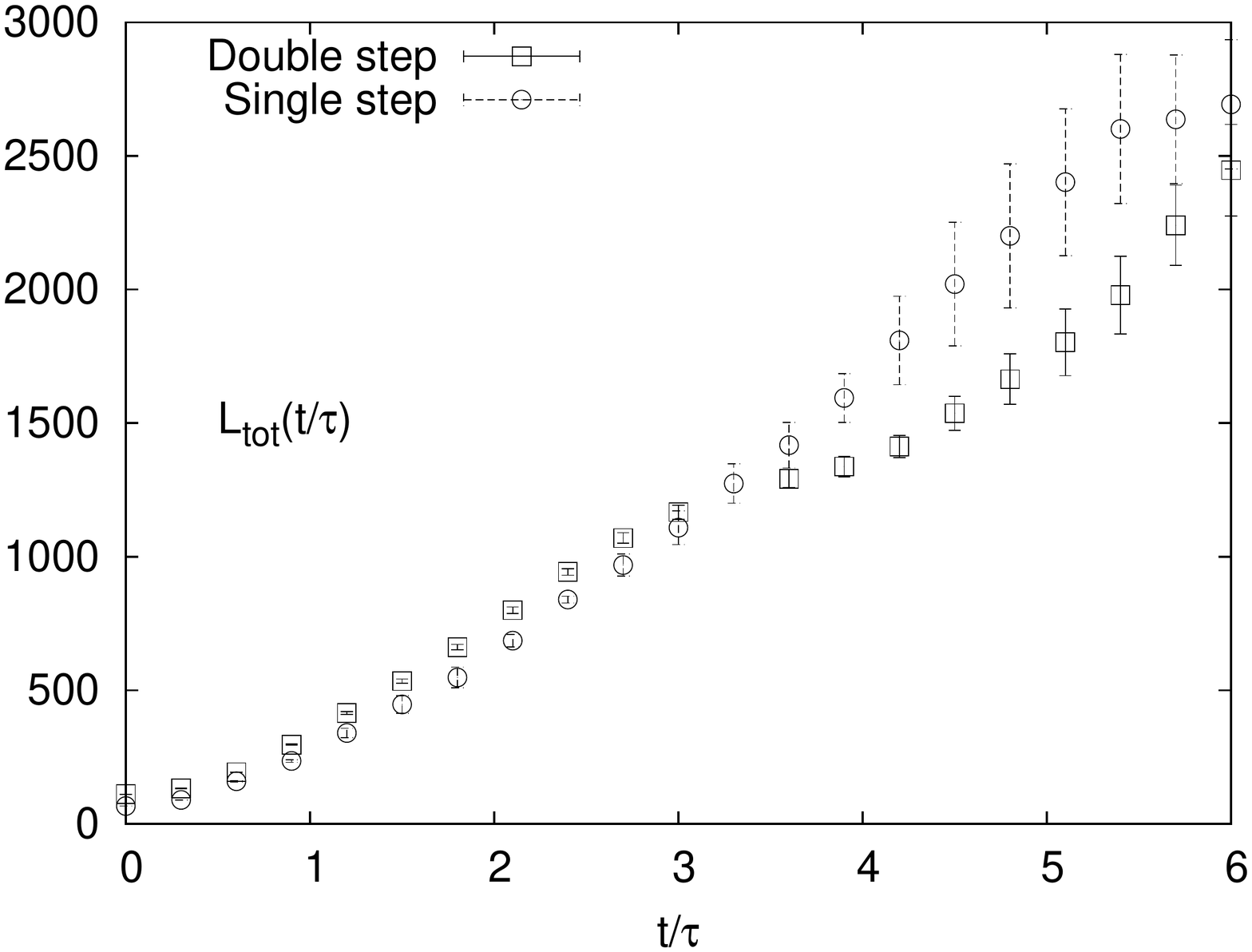}
\includegraphics[scale=0.35,angle=0]{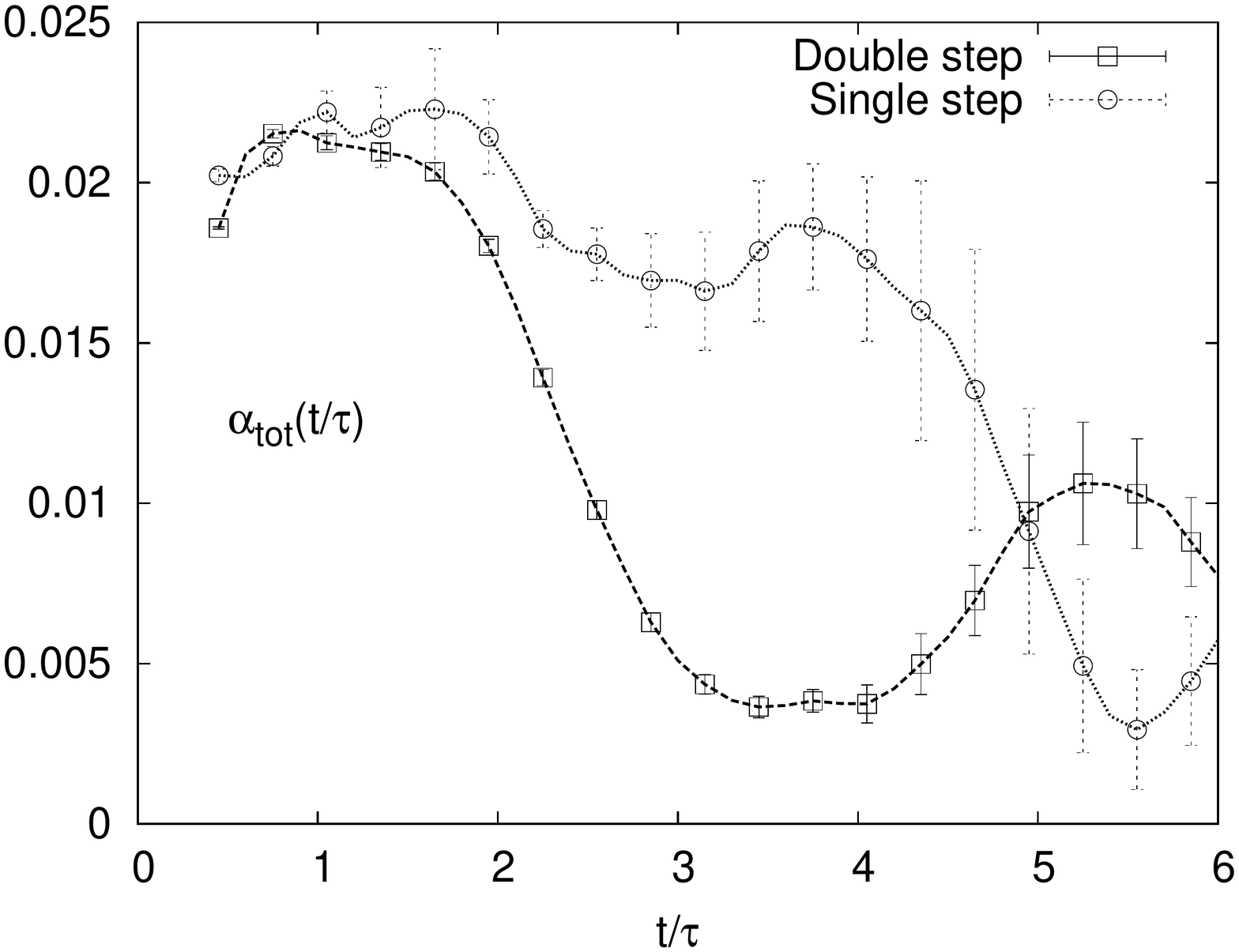}
\caption{Mixing layer length $L_{tot}(t/\tau)$ (left) and asymptotic growth rate $\alpha_{tot}(t/\tau)$ (right) evolution for the double (square) and single (circle) front RT.}
\label{mixing}
\end{figure}
\end{center}
In Figure \ref{mixing} we show the temporal evolution of the mixing length $L_{tot}(t)$ and the growth parameter $\alpha_{tot}(t)$ for the
 double-front and single-front RT. 
In order to make the analysis coherent, we apply the double map tent formulation (\ref{3L}) also to the single-front case, where now the middle temperature $T_m$ layer is ``virtual''. Inspection of Figure \ref{mixing} shows that --  as long as the two fronts evolve separately -- the mixing length has the same shape for the single and double-front cases  
with similar values of $\alpha_{tot}$, which is a good evidence that in this regime the two fronts proceed independently and in a self-similar way.
Differences arise after the two fronts come in contact and start to interact. As we can see the mixing rate of the double-RT decreases with respect to the single one, as it is 
visible from the drops of the curves  of $L_{tot}$ and $\alpha_{tot}$ in Figure \ref{mixing} around $t/\tau \sim 3$. Evidently, the turbulent viscosity generated from one front acts on the other slowing the propagation, as the system spends some amount of energy to
 fill the temperature/energy gap present between the two fronts (see Figure \ref{temperature} and Figure \ref{kinetic}  
for a one-to-one comparison of the temperature and kinetic energy fluctuations for single-front and double-front). For the classical RT system it is well known that temperature fluctuations remain constant during the development of convection (as we can see in the right panel of Figure \ref{temperature}); this is also valid for the double-front case until the two turbulent volumes 
 are separated, because we can treat each of them as an independent RT system. When they start to merge, the  temperature fluctuations in the region at the center of the cell, 
between the two fronts, must be ``re-ordered'' and brought to the same level of the two peaks, slowing the vertical growth of the mixing layer. After the two fronts are completely merged at $t/\tau{\sim}4$, the double-front system behaves like a classic one-front RT with temperatures $T_u$ and $T_d$, and the mixing layer length starts to increase again with  the expected rate, $L_{tot}\, {\sim}\, t^2$.
\begin{center}
\begin{figure}[h!]
\includegraphics[scale=0.35,angle=0]{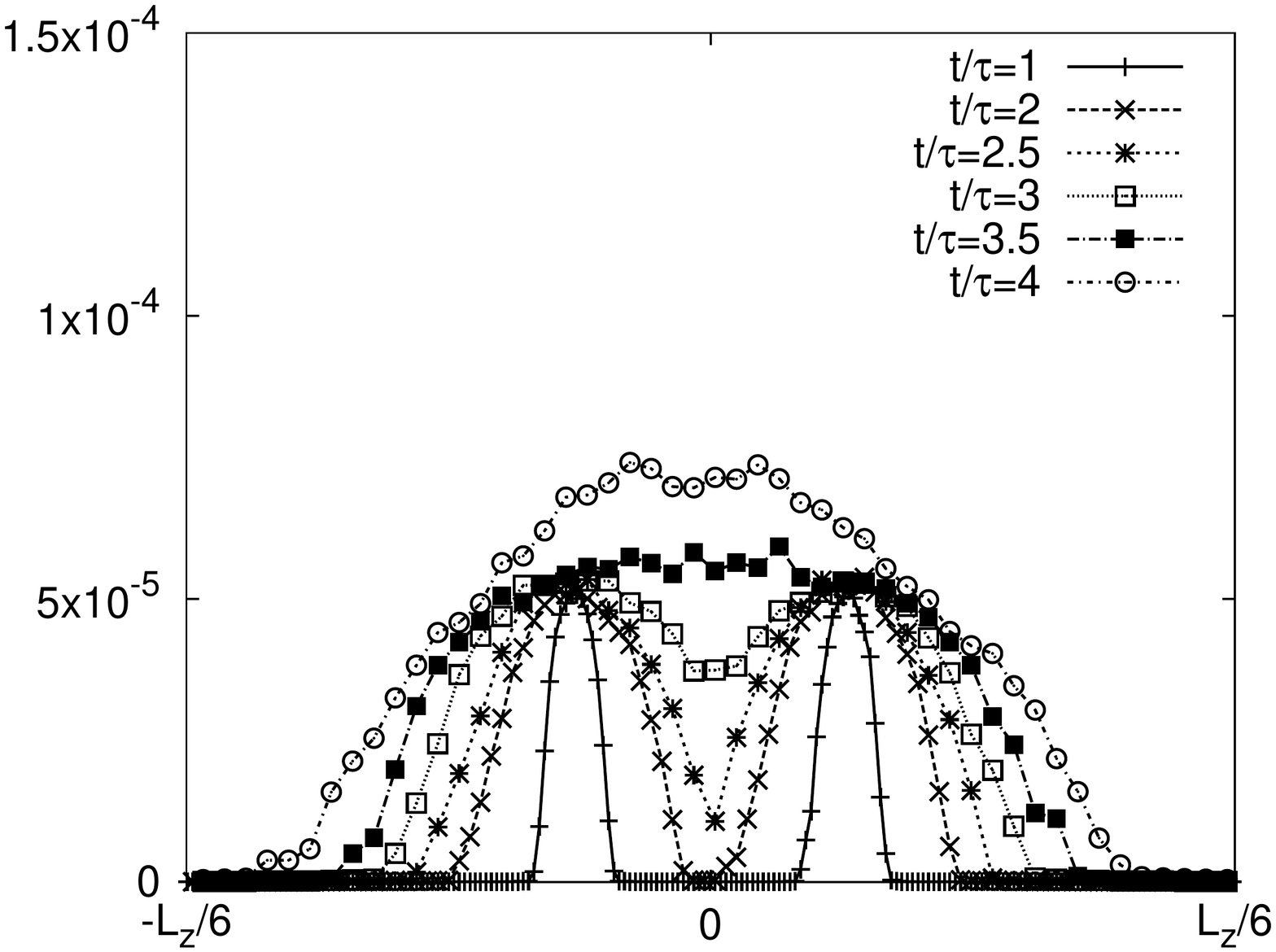}
\includegraphics[scale=0.35,angle=0]{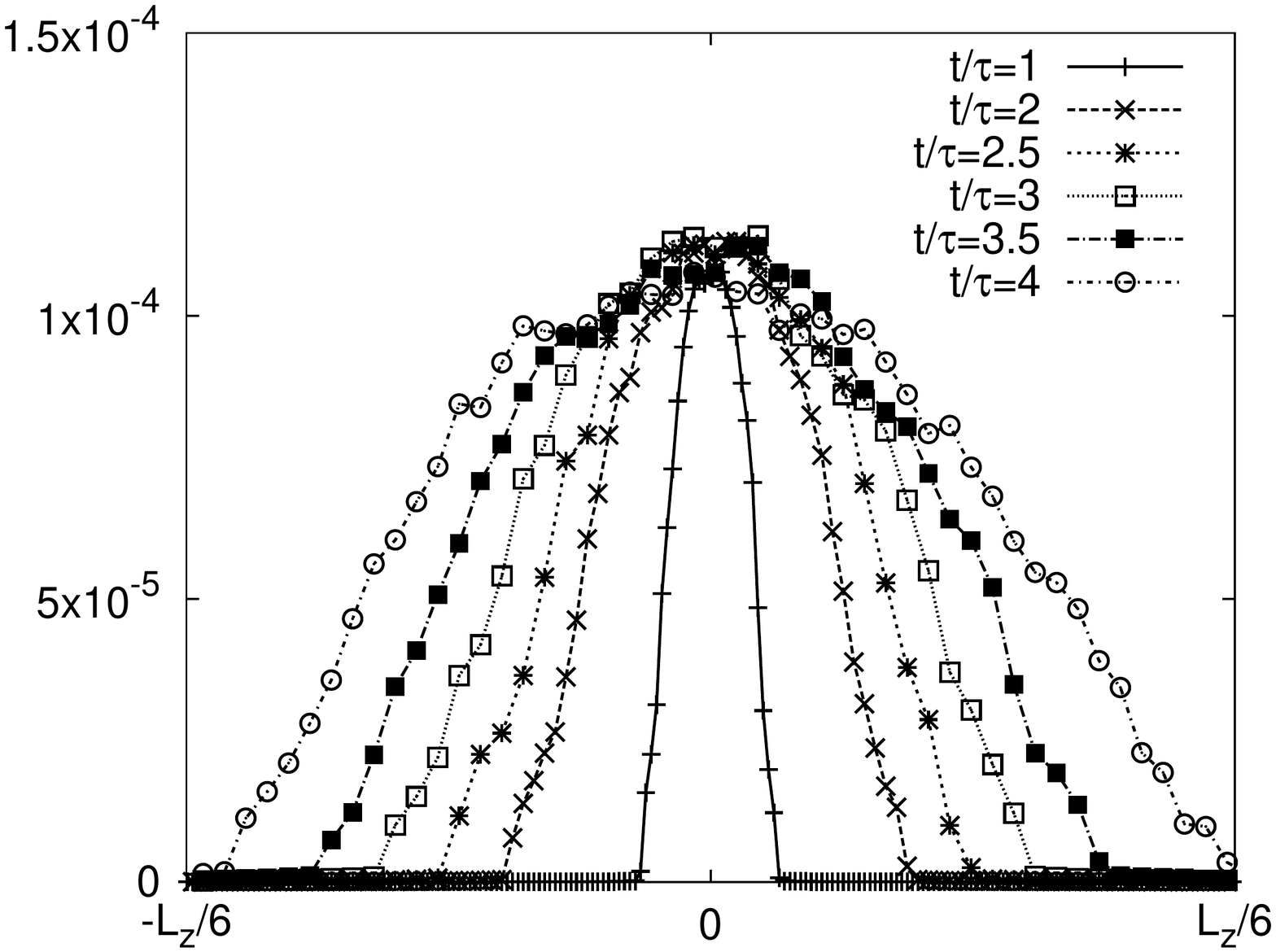}
\caption{Temperature fluctuations $\sqrt{\langle({T-\langle{T}\rangle_x})^2\rangle_x}$ for the double-front (right) and single-front (left) RT.}
\label{temperature}
\end{figure}
\end{center}
\begin{center}
\begin{figure}[h!]
\includegraphics[scale=0.35,angle=0]{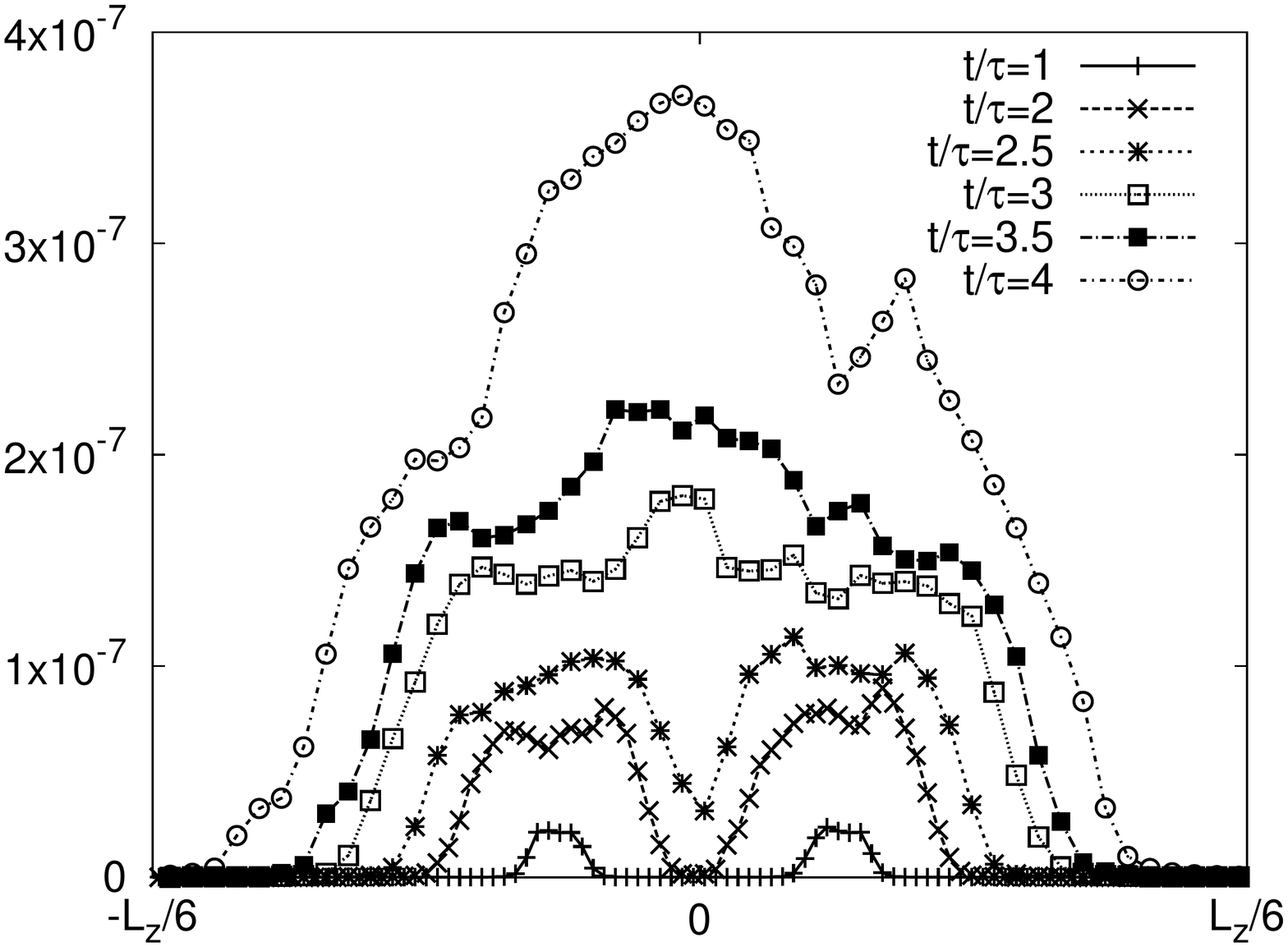}
\includegraphics[scale=0.35,angle=0]{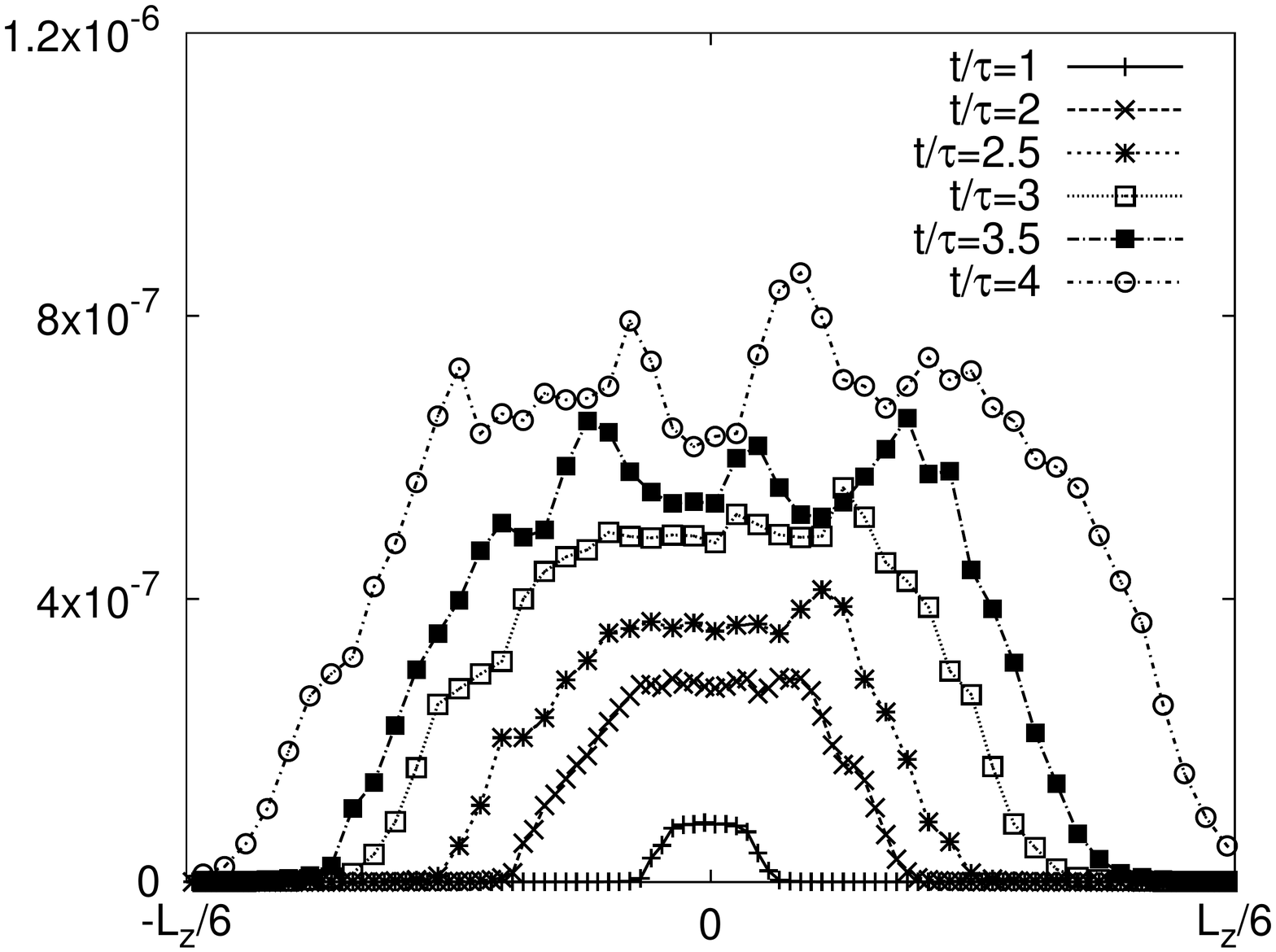}
\caption{Kinetic energy fluctuations $\sqrt{\langle({E_k-\langle{E_k}\rangle_x})^2\rangle_x}$ for the double-front (left) and single-front (right) RT.}
\label{kinetic}
\end{figure}
\end{center}
\section{Conclusions}
In this paper, we have studied a 2d Rayleigh-Taylor turbulence with an intermediate temperature layer 
in the middle of the vertical domain. The goal of this paper is twofold. First, we have developed a highly optimized GPU-based thermal Lattice Boltzmann algorithm to study 
hydrodynamical and thermal fluctuations in turbulent single phase fluids. Second, we have applied it to study the evolution (and collision) of two  turbulent RT fronts. 
Our results clearly show that at the moment of the collision the vertical evolution of the two front systems slows down and that the long time evolution is recovered only when a well mixed region is present in the center of the mixing length. This observation clearly shows the importance on the
outer environment on the evolution of any  RT system. Different initial temperature profiles would have lead to a different time for the collision between the two fronts and to a different duration of the intermediate mixing period (where the slow-down is observed). We have checked that the maximum effect is obtained in the configuration analyzed here, i.e. when two fronts of comparable kinetic energy collide. In the case when one of the two fronts is much stronger $At_u >> At_d$ the non-linear superposition becomes less important. 
Furthermore, a closer look at the spatial configurations for the one-front and two-front cases presented  in Figure \ref{figureT} demonstrates that the large-scale behavior recovers a universal evolution in the asymptotic regime while the temperature and velocity fluctuations at small-scales still  show important differences, suggesting the possibility of a long term memory of the initial configuration for high wave numbers modes. 
For completeness, we point out that the results presented here are related to the case of ``white noise'' initial conditions at the RT unstable interface with low values of the Atwood number. Differences can arise for the single-mode initial conditions and if the Atwood number is close to unity; in that case the nonlinear RT instability may be initially described as large-scale bubbles rising up in the heavy fluid with or without mixing, the latter depending on the strength of the secondary Kelvin-Helmotz instability \cite{Ramaprabhu,Oparin,Bychkov2}. A further generalization of this study to full 3d geometries and to analyze the small-scales properties of the system in the region where the two fronts collide will be presented in a future work. 

\section*{Acknowledgments}
We would like to thank CINECA (Bologna, Italy) for the use of their GPU-based computer resources, in the framework of the ISCRA Access Programme. This work has been supported by the SUMA project of INFN.  L.B. acknowledge partial funding from the European Research Council under the European Community's Seventh Framework Programme,  ERC Grant Agreement N. 339032.

\end{document}